\documentclass[aps,pre,twocolumn,superscriptaddress,nobibnotes,nodoi,noeprint]{revtex4-2}
\usepackage{amsmath,amssymb,physics,bm,siunitx}
\usepackage{xcolor,graphicx} \usepackage{soul}
\usepackage[colorlinks=true,citecolor=blue,urlcolor=blue,linkcolor=black]{hyperref}

\newcommand{\kB}{k_{\rm\scriptscriptstyle B}}
\newcommand{\rms}{\rm\scriptscriptstyle}

\newcommand{\rhoonemax}{\rho_1^{\rm max}}
\newcommand{\rhotwomax}{\rho_2^{\rm max}}  

\DeclareMathOperator*{\Vc}{V_{\rm c}}

\begin{document}

\title{Correlations of density and current fluctuations in single-file
  motion of hard spheres and in driven lattice gas with nearest-neighbor
  interaction}

\author{S\"{o}ren Schweers} \email{sschweers@uos.de}
\affiliation{Universit\"{a}t Osnabr\"{u}ck, Institute for Physics, 
 Barbarastra{\ss}e 7, D-49076
  Osnabr\"uck, Germany}

\author{Gunter M. Sch\"utz} \email{gschuetz04@yahoo.com}
\affiliation{IAS 2, Forschungszentrum J\"{u}lich, 52425 J\"{u}lich,
  Germany.}

\author{Philipp Maass} \email{maass@uos.de}
\affiliation{Universit\"{a}t Osnabr\"{u}ck, Institute for Physics, Barbarastra{\ss}e 7, D-49076
  Osnabr\"uck, Germany}

\date{February 20, 2025}

\begin{abstract}
We analyze correlations between density fluctuations and between
current fluctuations in a one-dimensional driven lattice gas with repulsive nearest-neighbor interaction and in
single-file Brownian motion of hard spheres dragged across a cosine potential with constant force.
By extensive  kinetic Monte Carlo and Brownian dynamics simulations we show that density and current correlation functions 
in nonequilibrium steady states follow the scaling behavior of the Kardar-Parisi-Zhang (KPZ) universality class. 
In a coordinate frame comoving with the collective particle velocity, 
the current correlation function decays as $\sim -t^{-4/3}$ with time $t$.
Density fluctuations spread superdiffusively as $\sim t^{2/3}$ at long times and
their spatio-temporal behavior is well described by the KPZ scaling function. 
In the absence of the cosine potential,  the correlation functions
in the system of dragged hard spheres show scaling behavior according to the Edwards-Wilkinson universality class.
In the coordinate frame comoving with the mean particle velocity,
they behave as in equilibrium, with
current correlations decaying as $\sim -t^{-3/2}$ and density fluctuations spreading
diffusively as $\sim t^{1/2}$.
\end{abstract}

\maketitle

\section{Introduction}

In single-file motion, particles cannot overtake each other and
accordingly keep their ordering \cite{Kaerger:2014, Nygard:2017,
  Taloni/etal:2017, Bukowski/etal:2021}.  This type of motion occurs,
for example, in confined colloidal systems \cite{Wei/etal:2000, Cui/etal:2002,
  Lin/etal:2002, Lutz/etal:2004a, Lutz/etal:2004b, Lin/etal:2005,
  Koeppl/etal:2006, Henseler/etal:2010}, zeolites
\cite{Hahn/etal:1996, Hahn/Kaerger:1998, Chmelik/etal:2018}, nanotubes
\cite{Cheng/Bowers:2007, Das/etal:2010, Dvoyashkin/etal:2014,
  Cao/etal:2018, Zeng/etal:2018}, membrane channels
\cite{Graf/etal:2000, Mamonov/etal:2003, Graf/etal:2004, Bauer/Nadler:2006, Kahms/etal:2009, Yang/etal:2010, Einax/etal:2010, Nelson:2011, Luan/Zhou:2018, Zhao/etal:2018, Kaerger/etal:2021}, 
  molecular motors
along filaments \cite{Lipowsky/etal:2010, Chowdhury:2013,
  Kolomeisky:2015, Jindal/etal:2020}, and ribosome translation
\cite{MacDonald/etal:1968, Schuetz:1997, Ciandrini/etal:2010, Klumpp/Hwa:2008, Zia/etal:2011, Erdmann-Pham:2020, Keisers/Krug:2023, Chowdhury/etal:2024}.  Particle
density fluctuations and current fluctuations in these systems are
intertwined \cite{Praehofer/Spohn:2002, Ferrari/Spohn:2016} and reflect
phenomena like superdiffusive spreading of density fluctuations
\cite{Beijeren/etal:1985, Popkov/etal:2015} and excess mass flow through interfaces
\cite{Beijeren:1991}. They are key for understanding subtle effects
in nonequilibrium dynamics, as, for example, recently found long-range
anticorrelations in the stochastic motion of domain walls separating
phases in driven diffusive systems \cite{Schweers/etal:2024}.

In equilibrium, autocorrelations of density and current fluctuations
take a Gaussian shape, while they are usually non-Gaussian in nonequilibrium
steady states (NESS) and can be described by the scaling function of
the Kardar-Parisi-Zhang (KPZ) universality class
\cite{Kardar/etal:1986, Ferrari/Spohn:2016, Spohn:2017, Spohn:2020,
  Quastel/Sarkar:2023}.  KPZ scaling is seen in many models, such as
height-height correlations of interface growth under ballistic
deposition \cite{Kardar/etal:1986, Ferrari/Spohn:2016, Spohn:2017},
spin–spin correlations of the Heisenberg chain \cite{DeNardis/etal:2022}, correlations of
locally conserved quantities (mass, momentum, energy) in
one-dimensional Newtonian ﬂuids \cite{Spohn:2020}, as well as density
correlations in driven lattice gases
\cite{Praehofer/Spohn:2002, Priyanka/Jain:2015, deGier/etal:2019}.

For diffusive systems, studies of density and current correlations
have so far focused on the symmetric and totally asymmetric simple
exclusion processes (SEP and TASEP) \cite{Derrida:1998, Schuetz:2001}. In these models, particles jump
between nearest sites of a one-dimensional lattice, where each lattice
site can be occupied by at most one particle, corresponding to an
exclusion interaction. In the SEP the rates for jumps to the left and
right are equal, while in the TASEP jumps are allowed in one direction
only.

In the SEP with mean particle density $\bar\rho$, correlation
functions
\begin{equation}
C_{\rho\rho}(x,t)=\langle
\delta\rho(x,t)\delta\rho(0,0)\rangle=\langle
\rho(x,t)\rho(0,0)\rangle-\bar\rho^2
\label{eq:autocorrelation_density_fluctuations}
\end{equation}
between fluctuations $\delta\rho(x,t)=\rho(x,t)-\bar\rho$ of the local
particle density $\rho(x,t)$ spread diffusively in the equilibrium
state, i.e.\ they broaden with the square root of time. They
obey the scaling law
\begin{equation}
C_{\rho\rho}(x,t)\sim\frac{\kappa}{\sqrt{2D_{\rho\rho}t}}\,G\left(\frac{x}{\sqrt{2D_{\rho\rho}t}}\right)
\label{eq:C-Gaussian-scaling}
\end{equation}
of the Edwards-Wilkinson (EW) universality class \cite{Krug:1997},
where $\kappa=\int\dd x\, C_{\rho\rho}(x,t)$, $D_{\rho\rho}$ is a
$\bar\rho$-dependent diffusion coefficient, and $G(.)$ is the standard
Gaussian, $G(u)=\exp(-u^2/2)/\sqrt{2\pi}$ \cite{Ferrari/Spohn:2016}.
Time correlations
\begin{equation}
C_{\scriptscriptstyle \!J\!J}(t)=\langle J(x,t) J(x,0)\rangle
\label{eq:autocorrelation_current}
\end{equation}
between currents $J(x,t)$ at the same position (site) $x$ decay as
\cite{Ferrari/Spohn:2016}
\begin{equation}
C_{\scriptscriptstyle \!J\!J}(t)\sim -A_{\rm eq}\, t^{-3/2}\,,\hspace{1em}t\to\infty\,,
\label{eq:current_correlations_equilibrium}
\end{equation}    
where $A_{\rm eq}$ is a constant with approximate value
\begin{equation}
A_{\rm eq}\simeq\frac{1}{8}D_{\rho\rho}^{1/2}\kappa\int\hspace{-0.2em} \dd x |x| G(x)=
\frac{\kappa}{4\sqrt{2\pi}}\,D_{\rho\rho}^{1/2}\,.
\label{eq:Aeq}
\end{equation}    

For the TASEP, density and current correlations show scaling behavior in a
nonequilibrium steady state (NESS), when they are recorded in a comoving
frame with collective velocity \cite{Schuetz:2001}
\begin{equation}
v=v(\bar\rho)=\frac{\dd \bar J(\bar\rho)}{\dd\bar\rho}\,,
\label{eq:v}
\end{equation}
where $\bar J(\bar\rho)$ is the stationary current at mean particle density
$\bar\rho$.
Density fluctuations spread superdiffusively in this frame, broadening
in time as $\sim t^{2/3}$.  The $C_{\rho\rho}(x,t)$ obey the scaling
form
\begin{equation}
C_{\rho\rho}(x,t)\sim\frac{\kappa}{(\Gamma_{\rho\rho} t)^{2/3}}\,F_{\rms
  KPZ}\left(\frac{x-vt}{(\Gamma_{\rho\rho} t)^{2/3}}\right)\,,
\label{eq:C-KPZ-scaling}
\end{equation}
where $\kappa=\int\dd x\,C_{\rho\rho}(x,t)$, $\Gamma_{\rho\rho}$ is a
$\bar\rho$-dependent rate of superdiffusive spreading, and $F_{\rms
  KPZ}(.)$ is the KPZ scaling function.  Its exact form was derived by
Pr\"ahofer and Spohn \cite{Praehofer/Spohn:2004}.
The rate of spreading is proportional to the absolute value  of the second derivative $\bar J''(\bar\rho)$
\cite{Krug/etal:1992, Ferrari/Spohn:2016}, $\Gamma_{\rho\rho}\propto |\bar J''(\bar\rho)|$.

The superdiffusive scaling of density fluctuations in NESS is
reflected in long-range anticorrelations between currents across an
interface moving with velocity $v(\bar\rho)$ \cite{Beijeren:1991}.
Considering the respective local currents $J_{\rm
  cf}(x,t)=J(x+vt,t)-v\rho(x+vt,t)$ in the comoving frame with average
$\bar J_{\rm cf}=\bar J-v\bar\rho$, 
the time correlations of their fluctuations
\begin{align}
C^{\rm cf}_{\scriptscriptstyle \!J\!J}(t)&= \langle\delta J_{\rm cf}(x,t)\delta J_{\rm cf}(x,0)\rangle\nonumber\\ 
&=\langle J_{\rm cf}(x,t)J_{\rm cf}(x,0)\rangle-\bar J_{\rm cf}^2
\label{eq:CjjKPZ}
\end{align}
decay as \cite{Ferrari/Spohn:2016}
\begin{equation}
C^{\rm cf}_{\scriptscriptstyle \!J\!J}(t)\sim -A\,t^{-4/3}\,,\hspace{1em}t\to\infty\,,
\label{eq:current_correlations_ness}
\end{equation}
where $A$ is a constant, which for zero collective velocity $v(\bar\rho)=0$ is approximately
\begin{equation}
A\simeq\frac{1}{9}\,\Gamma_{\rho\rho}^{2/3}\, \kappa \int\hspace{-0.2em} \dd x |x| F_{\rms KPZ}(x)\,.
\label{eq:A}
\end{equation}

Equations~\eqref{eq:autocorrelation_density_fluctuations}-\eqref{eq:A}
describe long-time asymptotic behavior and should hold true
irrespective of microscopic details for driven diffusive systems of
particles with short-range interactions. 
Here we test this for a
lattice gas with repulsive nearest-neighbor interactions and for a
system with continuous-space dynamics, given by overdamped Brownian
dynamics of hard spheres. In particular, we evaluate whether
correlations between current fluctuations decay with the expected power laws
and whether the long-time behavior becomes
visible beyond microscopic times scales without extended
transients. For the continuous-space system, we present a method to
cope with the problem of computing local fluctuations in a comoving
frame.
 
\section{Brownian motion of hard spheres and lattice gas with nearest-neighbor interaction}
\begin{figure}[t!]
\centering \includegraphics[width=\columnwidth]{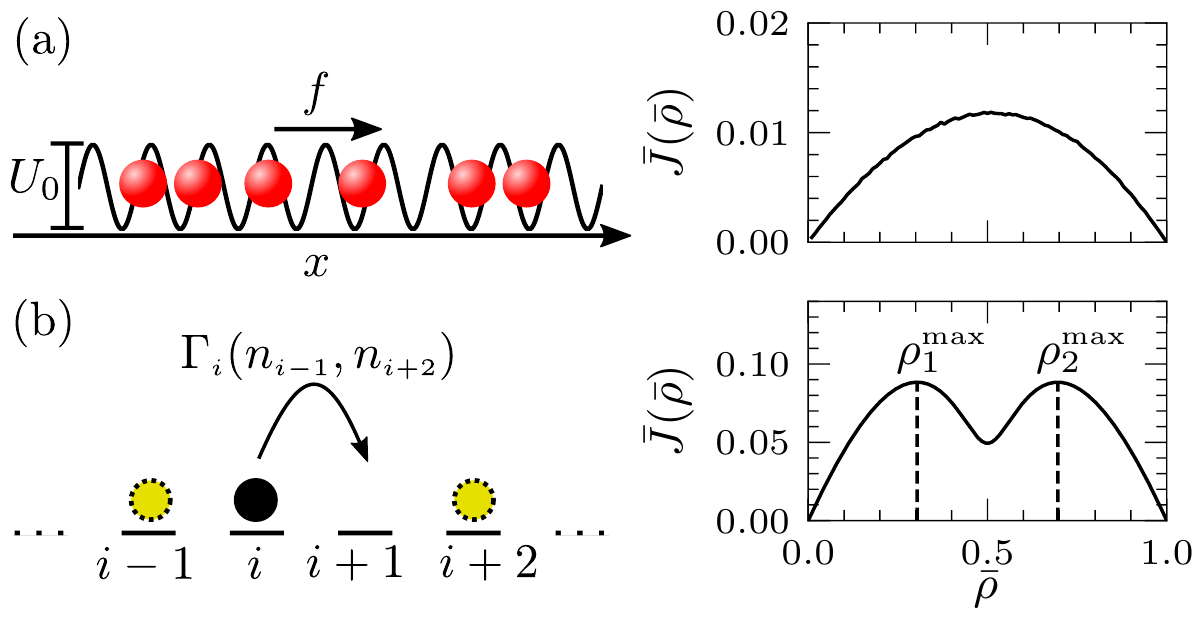}
\caption{(a) Left: Overdamped Brownian motion of hard spheres across a sinusoidal potential [Eq.~\eqref{eq:U(x)}] with barriers $U_0$ 
under a constant drag force $f$. Right:
Bulk current-density relation $\bar J(\bar\rho)$ for parameters $U_0/k_{\rm B}T=6$, $f\lambda/k_{\rm B}T=1$, and $\sigma=0.8\lambda$. (b) Left:
Particle hopping in a lattice with repulsive nearest-neighbor interactions, where rates $\Gamma(n_{i-1},n_{i+2})$ for a particle jump from a site $i$ to a vacant site $i+1$ depend on the occupation numbers $n_{i-1}$ and $n_{i+2}$ of neighboring sites $i-1$ and $i+2$. Right: Bulk current-density relation $\bar J(\bar\rho)$ for
Glauber rates [Eq.~\eqref{eq:glauber_rate}] with parameters $V/k_{\rm B}T=2V_c/k_{\rm B}T=4\ln 3$. It has two maxima at $\rhoonemax$ and $\rhotwomax$.}
\label{fig:models}
\end{figure}

Overdamped Brownian motion of hard spheres in a periodic potential and particle hopping 
in a lattice with repulsive nearest-neighbor interactions 
are illustrated in Fig.~\ref{fig:models} together with
current-density relations $\bar J(\bar\rho)$ in these systems. In the
following, we specify the dynamics in the two models.

\subsection{Brownian motion of hard spheres}
Figure~\ref{fig:models}(a) illustrates overdamped Brownian motion of $N$ hard spheres
with diameter $\sigma$ that are driven by a constant drag force $f$ across a sinusoidal potential
\begin{equation}
U(x)=\frac{U_0}{2}\cos(\frac{2\pi x}{\lambda})\,.
\label{eq:U(x)}
\end{equation}
Here, $\lambda$ is the wavelength of the potential and $U_0$ the barrier between potential wells.
This model was introduced as Brownian ASEP (BASEP), since it
can be viewed as extension of the ASEP to continuous-space space dynamics \cite{Lips/etal:2018}.

The BASEP constitutes a basis for understanding various intriguing phenomena occurring in Brownian single-file transport through
periodic potentials. This includes particle-size dependent current-density relations and 
non-equilibrium phase transition diagrams with up to five phases \cite{Lips/etal:2018, Lips/etal:2019, Lips/etal:2020}, 
faster uphill transitions in kinetics of local barrier passing
\cite{Ryabov/etal:2019}, effective barrier enhancements in flow-driven system caused by hydrodynamic
interactions \cite{Cereceda-Lopez/etal:2021, Lips/etal:2022},
ultrafast defect propagation and emergence of solitary cluster 
waves in dense systems \cite{Antonov/etal:2022b,  Antonov/etal:2022a, Cereceda-Lopez/etal:2023, Antonov/etal:2024}, as well as
phase synchronisation of collective dynamics and fractional Shapiro steps in particle currents under 
time-periodic driving \cite{Mishra/etal:2025}.

The particle motion in the BASEP is described by the Langevin equations
\begin{equation}
\frac{\dd x_i}{\dd t}=\mu \left(f-\frac{\partial U(x_i)}{\partial x}\right) + \sqrt{2D}\xi_i(t)\,,
\label{eq:langevin}
\end{equation}
where $x_i$, $i=1,\ldots,N$, are the particle positions, $\mu$ is the particle mobility,
$D=\kB T\mu$ is the diffusion constant, and  $\kB T$ thermal energy.
The $\xi_i(t)$ are uncorrelated stationary Gaussian processes with zero mean 
and correlation function $\langle\xi_i(t)\xi_j(t')\rangle=\delta_{ij}\delta(t-t')$.
Periodic boundary conditions are applied for a system of length $L$. The mean
particle density is $\bar\rho=N/L$.

The hard-sphere interaction implies the constraints
\begin{equation}
|x_i-x_j|\geq\sigma\,.
\end{equation}
Their consideration requires special care in simulations
\cite{Strating:1999,Scala/etal:2007,Scala:2012}. Here we use the
recently developed Brownian cluster dynamics algorithm
\cite{Antonov/etal:2022c, Antonov/etal:2025}.

The bulk current-density relation $\bar J(\bar\rho)$ for the BASEP
was studied in  \cite{Lips/etal:2018, Lips/etal:2019} and is shown
in Fig.~\ref{fig:models}(a) for $\sigma=0.8\lambda$, $U_0=6k_{\rm B}T$ and
$f=k_{\rm B}T/\lambda$.

For vanishing periodic potential ($U_0=0$), the current-density relation is
\begin{equation}
\bar J=\mu f\bar\rho\,.
\end{equation}
In this special case of a linear current-density relation, the EW rather than the KPZ
universality class applies: density and local current autocorrelations behave according to
Eqs.~\eqref{eq:C-Gaussian-scaling} and \eqref{eq:current_correlations_equilibrium}
as in an equilibrium system, if one determines the correlation functions in a frame comoving 
with the mean particle velocity $v(\bar\rho)=\dd\bar J(\bar\rho)/\dd\bar\rho=\mu f$. 
The non-applicability of KPZ scaling is reflected in 
$\Gamma_{\rho\rho}\propto |\bar J''(\bar\rho)|$ becoming zero for a linear current-density relation.

Physically, the absence of KPZ scaling for driven hard-spheres in a flat potential
$U_0=0$ can be understood from the fact that
dynamics in such system are equivalent to that of point particles \cite{Tonks:1938, Taloni/etal:2017}.
This follows after a simultaneous transformation $x_i\to x_i'=x_i-i\sigma$ of the particle coordinates and of the
system size, $L\to L'=L-N\sigma$.
For point particles, collective properties are the same as for independent particles \cite{Ryabov/Chvosta:2011}.
Accordingly, for such properties
the hard-sphere interaction becomes irrelevant in flat potentials. 
Let us note that this holds true also for dynamics of hard spheres
in the presence of an external periodic potential with wavelength $\lambda$, if the particle diameter is an integer multiple 
of $\lambda$ \cite{Lips/etal:2019}. 

We use $\lambda$, $k_{\rm B}T$, and $\lambda^2/D=\lambda^2/k_{\rm B}T\mu$ as
length, energy and time unit in the modeling; unit of force thus is $k_{\rm B}T/\lambda$.
If $U_0=0$, $\sigma$ replaces $\lambda$ as length unit.

\subsection{Lattice gas with nearest-neighbor interaction}
Figure~\ref{fig:models}(b) illustrates
particle hopping in a one-dimensional driven lattice gas 
with repulsive nearest-neighbor interaction $V$
\cite{Popkov/Schuetz:1999, Antal/Schuetz:2000, Dierl/etal:2012, Dierl/etal:2013}. 
The particles jump uni-directionally to unoccupied neighboring sites to their right.
The lattice has $L$ sites $i=1,\ldots,L$ and periodic boundary condition are applied.
A lattice site can be occupied by at most one particle.
Microstates of the system are given by the sets $\{n_i\}$ of occupation numbers, where
$n_i=1$ if site $i$ is occupied by a particle and zero otherwise.
The rate of a particle jump from site
$i$ to a vacant site $(i+1)$ is given by the Glauber form
\begin{equation}
\Gamma_i(n_{i-1},n_{i+2})=\frac{\nu}{\exp(\beta V[n_{i+2}-n_{i-1}])+1}\,,
\label{eq:glauber_rate}
\end{equation}
where $\nu$ is an attempt frequency and $\beta=1/k_{\rm B}T$.

We use the first-reaction time algorithm \cite{Gillespie:1978,Holubec/etal:2011} to simulate
hopping dynamics by kinetic Monte-Carlo simulations.

\begin{figure*}[t!]
\centering
\includegraphics[width=0.7\textwidth]{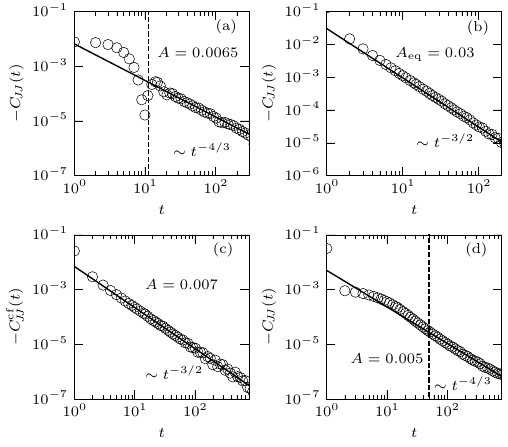}
\caption{Correlations between current fluctuations for (a) the TASEP with repulsive
next-neighbor interactions~$V=2\Vc$ at density $\bar\rho=\rhoonemax$ and (b)-(d) Brownian single-file motion of hard spheres 
with particle density $\bar\rho=1/2$. Data for hard-sphere systems are shown for three different situations:
(b) an equilibrium state in a flat potential
($U_0=0$), (c) a nonequilibrium steady state of dragged spheres ($f=1$)
in a flat potential ($U_0=0$), and (d) a nonequilibrium steady state of dragged spheres ($f=1$)
in a sinusoidal potential ($U_0=6$). Solid lines are least-square fits of the power laws with
exponents -3/2 [Eq.~\eqref{eq:current_correlations_equilibrium}] and -4/3 [Eq.~\eqref{eq:current_correlations_ness}]
to the asymptotic regime at long times (last 30 data points were used). The amplitude factors $A_{\rm eq}$ and $A$
are the parameters obtained from the fits.
Dashed vertical lines in (a) and (d) indicate estimates of times for the onset of the power law decay (see text).
Data points were obtained by averaging the correlation functions in small bins with equal spacing
on the logarithmic time axis.}
 \label{fig:Cjj}
\end{figure*}

The current-density relation is \cite{Popkov/Schuetz:1999, Dierl/etal:2012,Dierl/etal:2013}
\begin{equation}
\bar J(\bar\rho)= \nu
\left[\left(\bar\rho-C^{(1)}\right)^2\frac{2g-1}{2\bar\rho(1-\bar\rho)}+\left(\bar\rho-C^{(1)}\right)(1-g)\right]\label{eq:bulk_current_density_relation}
\end{equation}
where 
\begin{multline}
C^{(1)}=\frac{1}{2\left[1-\exp(-\beta V)\right]}\left[2\bar\rho(1-\exp(-\beta V))
-1\vphantom{\sqrt{1-4\bar\rho(1-\bar\rho)\left[1-\exp(-\beta V)\right]}}\right.\\\left.
+\sqrt{1-4\bar\rho(1-\bar\rho)\left[1-\exp(-\beta V)\right]}\right]
\label{eq:C1}
\end{multline}
and
\begin{equation}
g = \frac{1}{\exp(\beta V)+1}\,.
\end{equation}
$C^{(1)}$ in Eq.~\eqref{eq:C1}
is the correlation $\langle n_in_{i+1}\rangle_{\rm eq}$ in an equilibrium state of a lattice gas with nearest-neighbor interaction $V$.
If the interaction exceeds a critical value $\Vc$, $\beta V_c=2\ln3$, the
current-density relation has two degenerate maxima, 
$\bar J(\rhoonemax)=\bar J(\rhotwomax)$ with
$\rhoonemax\neq\rhotwomax$. For $V=2\Vc$, which we use in the following, these maxima are at
$\rhoonemax = (1-[3-\sqrt{8e^{\beta V}/(e^{\beta V}-1)}]^{1/2})/2\cong0.304$ and
$\rhotwomax=1-\rhoonemax\cong0.696$
\cite{Dierl/etal:2012,Dierl/etal:2013}.

We use the lattice constant $a$ and the inverse attempt frequency $\nu^{-1}$ as length and time unit in the modeling of the driven lattice gas.

\section{Scaling of density correlations and long-range anticorrelations between current fluctuations}
\label{sec:results}
In the analysis of the hard sphere system,
we divide the system into 
$L/\sigma$ boxes of size $\sigma$, which can accommodate at most one particle. 
To each box $i$ at time $t$, we assign an occupation number 
$n_i(t)\in\{0,1\}$, i.e.\ we obtain a description as in a lattice gas with lattice constant $\sigma$.
This way we can calculate correlation functions as for a lattice gas.

To determine $C_{\rho\rho}(x,t)$,
we first generate time series of the occupation numbers $n_i(k\Delta t)$, $k=1,\ldots, M$,
with time step $\Delta t=\nu^{-1}$ in the lattice gas and time step $\Delta t=\sigma^2/D$
in the hard-sphere system. In equilibrium and nonequilibrium steady states, we then correlate 
fluctuations $\delta n_i(k'\Delta t)=n_i(k'\Delta t)-\bar n$ and $\delta n_{i+j}(k'\Delta t+k\Delta t)$
at distance $x=ja$ and time lag $t=k\Delta t$.

For calculating current correlations in the comoving frame, 
we determine link currents
$j_{i,i+1}(t)=[n(i\to i\!+\!1)-n(i\!+\!1\to i)]/\Delta t$ between all pairs of neighboring lattice sites $i$, $i+1$ 
at discrete times $t=l\Delta t$, $l=1,2\ldots$, where
$n(i\to i\!+\!1)$ and $n(i\!+\!1\to i)$ are the numbers of particles moving from site $i$ to $(i+1)$ and vice versa in
the time interval $]t,t+\Delta t]$. The link currents $j^{\rm cf}_{i,i+1}(t)$ at time $t$ in the 
frame comoving with the collective velocity $v$ [Eq.~\eqref{eq:v}] are
$j^{\rm cf}_{i,i+1}(t)=j_{k,k+1}(t)-n_{k+1}(t+\Delta t)v$, where $k=j+vt$.
For $v\ne0$, we set $\Delta t=1/|v|$ for the time step $\Delta t$, guaranteeing
that $vt$ is an integer. For $v=0$, we use the same $\Delta t$ as in the calculation of $C_{\rho\rho}(x,t)$.

\subsection{Long-range anticorrelations between local current fluctuations}
\label{subsec:current_fluctuations}
Figure~\ref{fig:Cjj} shows correlation functions of local current fluctuations 
for (a) the driven lattice gas at $\bar\rho=\rhoonemax$ [see Fig.~\ref{fig:models}(b)], 
and (b)-(d) the hard sphere system at $\bar\rho=1/2$.
In Fig.~\ref{fig:Cjj}(b) the results are for the equilibrium system in the absence of the periodic potential, i.e.\
for $f=0$ and $U_0=0$. 
In Figs.~\ref{fig:Cjj}(c) and (d) the results are for the nonequilibrium system ($f=1$),
where in (c) $U_0=0$, and in (d) $U_0=6$ and $\sigma=0.8$. 
As the current fluctuations are anticorrelated, we plotted
the negative of the correlation functions in the double-logarithmic representations. 
In Fig.~\ref{fig:Cjj}(c), the collective velocity is $v=\mu f=1$, and it is zero in Figs.~\ref{fig:Cjj}(a),(b) and (d).

In all cases the current correlation functions show the expected power law decay at large times:
$C_{\scriptscriptstyle \!J\!J}(t)\sim t^{-4/3}$ in Figs.~\ref{fig:Cjj}(a),(d), and a decay $\sim t^{-3/2}$ 
in Figs.~\ref{fig:Cjj}(b),(c).

\begin{figure*}[t!]
\centering
\includegraphics[width=\textwidth]{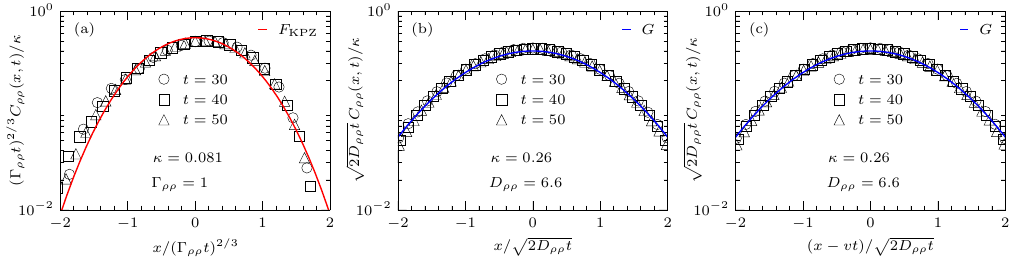}
\caption{Scaled correlation functions of particle density fluctuations for the same systems as in Figs.~\ref{fig:Cjj}(a)-(c).
The $\kappa$ values were obtained by numerical integration of the unscaled correlation functions $C_{\rho\rho}(x,t)$. Parameters
$\Gamma_{\rho\rho}$ and $D_{\rho\rho}$ were found by an optimal match of the normalized
functions $F_{\rms KPZ}(.)$ and $G(.)$ in Eqs.~\eqref{eq:C-KPZ-scaling} and \eqref{eq:C-Gaussian-scaling} to the scaled simulated data.}
\label{fig:Crhorho-scaled}
\end{figure*}

Amplitude factors $A_{\rm eq}$ and $A$ of the power laws were obtained from a least-square fitting of Eqs.~\eqref{eq:current_correlations_equilibrium}
and \eqref{eq:current_correlations_ness} to the long-time regimes of the simulated data, see the
values given in Figs~\ref{fig:Cjj}(a)-(d). For the situation considered in Fig.~\ref{fig:Cjj}(a) and (b), we 
can compare these
numbers with the predicted approximate values in Eqs.~\eqref{eq:Aeq} and \eqref{eq:A}.
To calculate $A$  and $A_{\rm eq}$
from these equations, we used the values $\kappa$, $\Gamma_{\rho\rho}$ or $D_{\rho\rho}$ obtained from
the analysis of the  
density autocorrelation functions in Sec.~\ref{subsec:density_fluctuations}, see Figs.~\ref{fig:Crhorho-scaled}(a)-(b).
The resulting $A\cong0.0051$ and $A_{\rm eq}\cong0.067$ are of the order as the values 
given in Figs.~\ref{fig:Cjj}(a) and (b).

In Figs.~\ref{fig:Cjj}(b),(c) the data approach the power law decay $~\sim t^{-3/2}$ from the beginning, while
a different behavior is seen in Figs.~\ref{fig:Cjj}(a),(d) at small microscopic time scales. 
The power law decay $\sim t^{-4/3}$ in Fig.~\ref{fig:Cjj}(a) sets in for times larger than 
the inverse mean hopping rate of the particle. In Fig.~\ref{fig:Cjj}(d),
the onset time can be estimated by the time 
for the particles to pass their typical spacing $1/\bar\rho-\sigma$. With the mean particle velocity $\bar J/\bar\rho$, we get 
$t_\times\simeq (1-\bar\rho\sigma)/\bar J$. Both estimated
onset times are indicated by the vertical dashed lines in Figs.~\ref{fig:Cjj}(a) and (d).

\subsection{Scaling behavior of correlations between density fluctuations}
\label{subsec:density_fluctuations}
While it is straightforward to determine density correlations in space and time by the numerical 
procedure described at the beginning of this section, a
thorough analysis of their scaling behavior is extremely costly regarding computing time 
and data storage requirements \footnote{For generating one data set, 1000 {CPU} 
cores of a HPC cluster with {AMD} {EPYC} 7742 {CPUs} were used for one day.}. We could not obtain
$C_{\rho\rho}(x,t)$ over many orders of magnitudes in space and time, which would be necessary 
to check the theoretically predicted scaling properties
based on the simulated data for $C_{\rho\rho}(x,t)$ alone. However, using the results 
from Sec.~\ref{subsec:current_fluctuations} for the current correlations $C_{\scriptscriptstyle \!J\!J}(t)$, 
we can perform an analysis with the assurance that scaling behavior is  either according to Eq.~\eqref{eq:C-Gaussian-scaling} 
or Eq.~\eqref{eq:C-KPZ-scaling}.

Figures~\ref{fig:Crhorho-scaled}(a)-(c) show scaling plots of $C_{\rho\rho}(x,t)$ 
according to Eqs.~\eqref{eq:C-Gaussian-scaling} and \eqref{eq:C-KPZ-scaling}
for the same systems as in Figs.~\ref{fig:Cjj}(a)-(c).
For the hard-spheres driven across the sinusoidal potential, we could not obtain data with sufficient numerical accuracy 
over a significant length scale covering several wavelengths $\lambda$. 

The scaled data in Figs.~\ref{fig:Crhorho-scaled}(a)-(c) for three different times collapse onto common master curves.
We determined the constant $\kappa$ by numerical integration of $C_{\rho\rho}(x,t)$. For obtaining the value $\Gamma_{\rho\rho}=1$ in 
Fig.~\ref{fig:Crhorho-scaled}(a)  we fitted the parameter $\Gamma_{\rho\rho}$ by an 
optimal match of $(\Gamma_{\rho\rho}t)^{2/3}C_{\rho\rho}(x,t)/\kappa$ to the KPZ scaling function. 
For $F_{\rms KPZ}(.)$, we used the tabulated data in \cite{Praehofer/SpohnKPZfunction:2002}. 
Likewise, the value $D_{\rho\rho}=6.6$ in Figs.~\ref{fig:Crhorho-scaled}(b),(c) was
obtained by fitting the parameter $D_{\rho\rho}$ to give an optimal match with the standard Gaussian. 

\section{Summary and Conclusions}
We have tested the appearance of EW and KPZ scaling for correlations between
density fluctuations and for related power law decays of correlations between
local current fluctuations in single-file dynamics of a discrete and continuous-space systems.
For the discrete system, we considered a driven lattice gas with repulsive nearest-neighbor interactions.
For a representative system with continuous-space dynamics we studied Brownian motion of hard spheres in different situations: 
equilibrium dynamics in a flat potential as well as driven transport under a constant drag force in a flat and periodic potential 
with barriers of height six times the thermal energy.

Our simulation results are in agreement with theoretically predicted scaling behaviors: for
the driven lattice gas and the hard spheres driven across the periodic potential, local current correlations
are anticorrelated at long times and decay as $\sim t^{-4/3}$. Density correlations for the lattice gas spread superdiffusively 
as $\sim t^{2/3}$ and are described by the KPZ scaling function.
For hard spheres in the periodic potential,
a corresponding analysis of correlations between density fluctuations was unfortunately not feasible with our computational resources.

In the flat potential, the single-file dynamics of hard spheres showed scaling features according to the EW universality class 
for both the equilibrium and the driven system: correlations between local current fluctuations 
are anticorrelated and decay as $\sim t^{-3/2}$ at long times and
correlations between density fluctuations spread diffusively and are described by the Gaussian scaling function. For the driven system,
Gaussian scaling is present because the current-density relation is linear.

Long-range anticorrelated current fluctuations in driven single file transport are important for stochastic dynamics
of domain walls between coexisting nonequilibrium phases, which occur in open systems coupled to particle reservoirs \cite{Krug:1991, Popkov/Schuetz:1999}. 
As shown recently \cite{Schweers/etal:2024}, the power-law decay $\sim t^{-4/3}$ 
leads to a $t^{1/3}$ subdiffusive growth of the width 
of domain wall position fluctuations, if the domain walls 
separate coexisting extremal
current phases. 
The collective velocity $v(\bar\rho)=\dd\bar J(\bar\rho)/\dd\bar\rho$ is
zero in both coexisting  phases in that case because the densities of these phases are extrema of the
bulk current-density relation $\bar J(\bar\rho)$. 
Generically, for domain walls separating coexisting phases 
the width of wall position fluctuations grows diffusively as
$t^{1/2}$ as long as $t^{1/2}/L \ll 1 $ in a system
of length $L$ \cite{Schuetz/Domany:1993, Derrida/etal:1993, Kolomeisky/etal:1998, Dudzinski/Schuetz:2000}.
This scaling behavior corresponds to an effective random walk dynamics of the domain wall position.
Indeed, this random walk property, which implies a Gaussian shape of the distribution
of the domain wall position, has recently been proved rigorously \cite{Schuetz:2023} for some parameter manifolds
in the ASEP. An intriguing open question is the distribution of the position in the case
of the subdiffusive growth of the position fluctuations \cite{Schweers/etal:2024}.

Mapping of particle positions in continuous-space
systems to occupation numbers of small bins is a suitable method also for testing scaling behavior of density correlations
in experiments.
For the interpretation of experimental data, one can estimate $v(\bar\rho)=\dd\bar J(\bar\rho)/\dd\bar\rho$
by measuring particle currents at densities around $\bar\rho$
and by taking a numerical derivative. 

\begin{acknowledgments}
This work has been funded by the Deutsche Forschungsgemeinschaft (DFG,
Project No.\ 355031190), by FCT/Portugal
through project UIDB/04459/2020 with DOI identifier 10-54499/UIDP/04459/2020, 
and through the grant 2022.09232.PTDC
with DOI identifier 10.54499/2022.09232.PTDC. We sincerely thank A.~Ryabov and the members of the DFG Research
Unit FOR2692 for fruitful discussions. We acknowledge use of a
high-performance computing cluster funded by the Deutsche
Forschungsgemeinschaft (DFG, Project No.\ 456666331).
\end{acknowledgments}


%

\end{document}